\documentclass{aip-cp}

\usepackage[numbers]{natbib}
\usepackage{rotating}
\usepackage{graphicx}
\usepackage{url}
\usepackage{threeparttable}
\usepackage{wrapfig}

\begin{document}

\title{TARGET: A Digitizing And Trigger ASIC For The Cherenkov Telescope Array}

\author[aff1]{S.~Funk}
\author[aff1]{D.~Jankowsky\corref{cor1}}
\author[aff2]{H.~Katagiri}
\author[aff1]{M.~Kraus\corref{cor2}}
\author[aff3]{A.~Okumura}
\author[aff4]{H.~Schoorlemmer}
\author[aff2]{A.~Shigenaka}
\author[aff3]{H.~Tajima}
\author[aff4]{L.~Tibaldo}
\author[aff5]{G.~Varner}
\author[aff1]{A.~Zink}
\author[aff4]{J.~Zorn}
\author{the CTA consortium\corref{cor3}}
\affil[aff1]{Erlangen Centre for Astroparticle Physics (ECAP), Erwin- Rommel-Str. 1, D 91058 Erlangen, Germany}
\affil[aff2]{College of Science, Ibaraki University, 2-1-1, Bunkyo, Mito 310-8512, Japan}
\affil[aff3]{Institute for Space-Earth Environmental Research, Nagoya University, Furo-cho, Chikusa-ku, Nagoya, Aichi 464-8601, Japan}
\affil[aff4]{Max-Planck-Institut f\"{u}r Kernphysik, P.O. Box 103980, D 69029 Heidelberg, Germany}
\affil[aff5]{Department of Physics and Astronomy, University of Hawaii, 2505 Correa Road, Honolulu, HI 96822, USA}

\corresp[cor1]{Corresponding author: david.jankowsky@fau.de}
\corresp[cor2]{Corresponding author: manuel.kraus@fau.de}
\corresp[cor3]{Full consortium author list at \url{http://cta-observatory.org}}

\maketitle

\begin{abstract}
The future ground-based gamma-ray observatory Cherenkov Telescope Array (CTA) will feature multiple types of imaging atmospheric Cherenkov telescopes, each with thousands of pixels. To be affordable, camera concepts for these telescopes have to feature low cost per channel and at the same time meet the requirements for CTA in order to achieve the desired scientific goals. We present the concept of the TeV Array Readout Electronics with GSa/s sampling and Event Trigger (TARGET) Application Specific Circuit (ASIC), envisaged to be used in the cameras of various CTA telescopes, e.g. the Gamma-ray Cherenkov Telescope (GCT), a proposed 2-Mirror Small-Sized Telescope, and the Schwarzschild-Couder Telescope (SCT), a proposed Medium-Sized Telescope. In the latest version of this readout concept the sampling and trigger parts are split into dedicated ASICs, TARGET~C and T5TEA, both providing 16 parallel input channels. TARGET C features a tunable sampling rate (usually 1 GSa/s), a 16k sample deep buffer for each channel and on-demand digitization and transmission of waveforms with typical spans of $\sim$100 ns. The trigger ASIC, T5TEA, provides 4 low voltage differential signal (LVDS) trigger outputs and can generate a pedestal voltage independently for each channel. Trigger signals are generated by T5TEA based on the analog sum of the input in four independent groups of four adjacent channels and compared to a threshold set by the user. Thus, T5TEA generates four LVDS trigger outputs, as well as 16 pedestal voltages fed to TARGET C independently for each channel. We show preliminary results of the characterization and testing of TARGET C and T5TEA.
\end{abstract}

\section{INTRODUCTION}

The TARGET application-specific-integrated circuit (ASIC) series has been designed and optimized specifically for the readout of Cherenkov cameras \cite{target}, such as the GCT \cite{GCT} (or see paper of L.~Tibaldo et al. in these proceedings) and SCT \cite{SCT} telescopes proposed for the Cherenkov Telescope Array (CTA) \cite{CTA}. CTA will feature a number of telescopes an order of magnitude higher than present arrays like H.E.S.S. or VERITAS and, therefore, the costs for the readout electronics are an important factor and should be reasonably low. Nevertheless, performance requirements given by the CTA consortium have to be met in order to reach the scientific goals aimed for. The proposed layout of the TARGET modules is capable of reading out 64 photodetector pixels and consists out of 8 TARGET ASICs and a companion field-programmable gate array (FPGA) and, thus, features a very limited number of components, making it both cheap and reliable.

The key features of all generations of TARGET ASICs are a tunable sampling frequency $\geq$ 1 GSa/s, a several $\mu$s deep analog buffer, a dynamic range $\geq$ 10 bits, trigger with adjustable threshold, a continuous sampling and almost deadtime free operation as well as the possibility to record full waveforms.

In order to minimize the crosstalk between the sampling and trigger paths, which occurred in former versions of TARGET, the two functions have been split into separate ASICs in the newest generation of TARGET ASICs, with the T5TEA ASIC for triggering and the TARGET~C ASIC for sampling and digitization. This step significantly improves the trigger performance whilst retaining the charge resolution and dynamic range of TARGET 7 \cite{ICRC2015}.

In this paper we give a short overview of the architecture of TARGET and present some measurements carried out to characterize the performance of the ASICs. We conclude with an outlook of future developments.

\section{TARGET C AND T5TEA ARCHITECTURE}

The main functional blocks of the ASICs are triggering (T5TEA), analog sampling and storage buffers, analog-to-digital converters (ADC) for digitizing the signals (TARGET C) and internal digital-to-analog converters (DAC) for setting operation values (both). The sampling and storage of signals is done using switched capacitor arrays (SCAs) where the signal is connected sequentially to the capacitors via analog switches. The sampling array consists of two blocks with 32 cells (capacitors) operated in ping pong mode: while data sampling is performed by one block, the other block is buffered to the storage array. In contrast, the storage array consists of a larger number of cells (16348 per channel) to guarantee a large buffer depth ($\sim 16$ $\mu$s). Wilkinson ADCs are used for digitizing the analog signals in the storage array on demand. A Wilkinson ramp generator passes an adjustable ramp to all channels and simultaneously a 12 bit counter starts counting until the ramp signal crosses the analog voltage. The measured counts then correspond to the value of the applied signal. The digitization is initiated by a request from the FPGA based on external or internal (i.e. from T5TEA) signals.
These features and other important ones (including performance results which will be explained in the following sections) for TARGET C and T5TEA are listed in Table~\ref{tab:targets}. For comparison we also show the performance of previous generations of TARGET, namely TARGET 5 and TARGET 7.

\begin{table}[htbp]
\begin{threeparttable}
\caption{Characteristics and performance of TARGET C + T5TEA compared to the former generations TARGET 5 and TARGET 7.}
\label{tab:a}
\tabcolsep7pt\begin{tabular}{llll}
\hline
  & \tch{1}{c}{b}{TARGET 5}  & \tch{1}{c}{b}{TARGET 7}  & \tch{1}{c}{b}{TARGET C \\+ T5TEA}   \\
\hline
\tch{1}{c}{b}{Characteristics} & & & \\
\hline
Number of Channels & 16 & 16 & 16 \\
Sampling frequency (Gsa/s) & 0.4 - 1 & 0.5 - 1 & 0.5 - 1 \\
Size of storage array & 16384 & 16384 & 16384 \\
Digitization clock speed (MHz) & $\sim$ 700 & 208 & 500 \\
Samples digitized simultaneously & 32 $\times$ 16 & 32 $\times$ 16 & 32 $\times$ 16 \\
Trigger (sum of 4 channels) & integrated & integrated & companion \\
\hline
\tch{1}{c}{b}{Performance} &  &  &  \\
\hline
Dynamic Range (V) & 1.1 & 1.9 & $\geq$ 1.9 \\
Integrated non linearity (mV) & 75 & 40 & $\leq$ 70 \\
Charge linearity range (pe\tnote{$\ast$} ) & 4 - 300 & 1 - $\geq$ 300 & 1 - $\geq$ 300 \\
Charge resolution at 10 pe\tnote{$\ast$} & 8\% & 4\% & $\leq$ 4\% \\
Charge resolution at $>$ 100 pe\tnote{$\ast$} & 2\% & $\leq$ 0.8\% & $\leq$ 0.8\% \\
Minimum trigger threshold (mV) & 20 & 50 & $\leq$ 8 \\
Trigger noise (mV) & 5 & 15 & $\leq$ 1 \\
\hline 
\end{tabular}
\label{tab:targets}
\begin{tablenotes}
\item [$\ast$] assuming 4 mV/pe
\end{tablenotes}
\end{threeparttable}
\end{table}

\begin{figure}[htbp]
  \centerline{\includegraphics[width=0.5\textwidth]{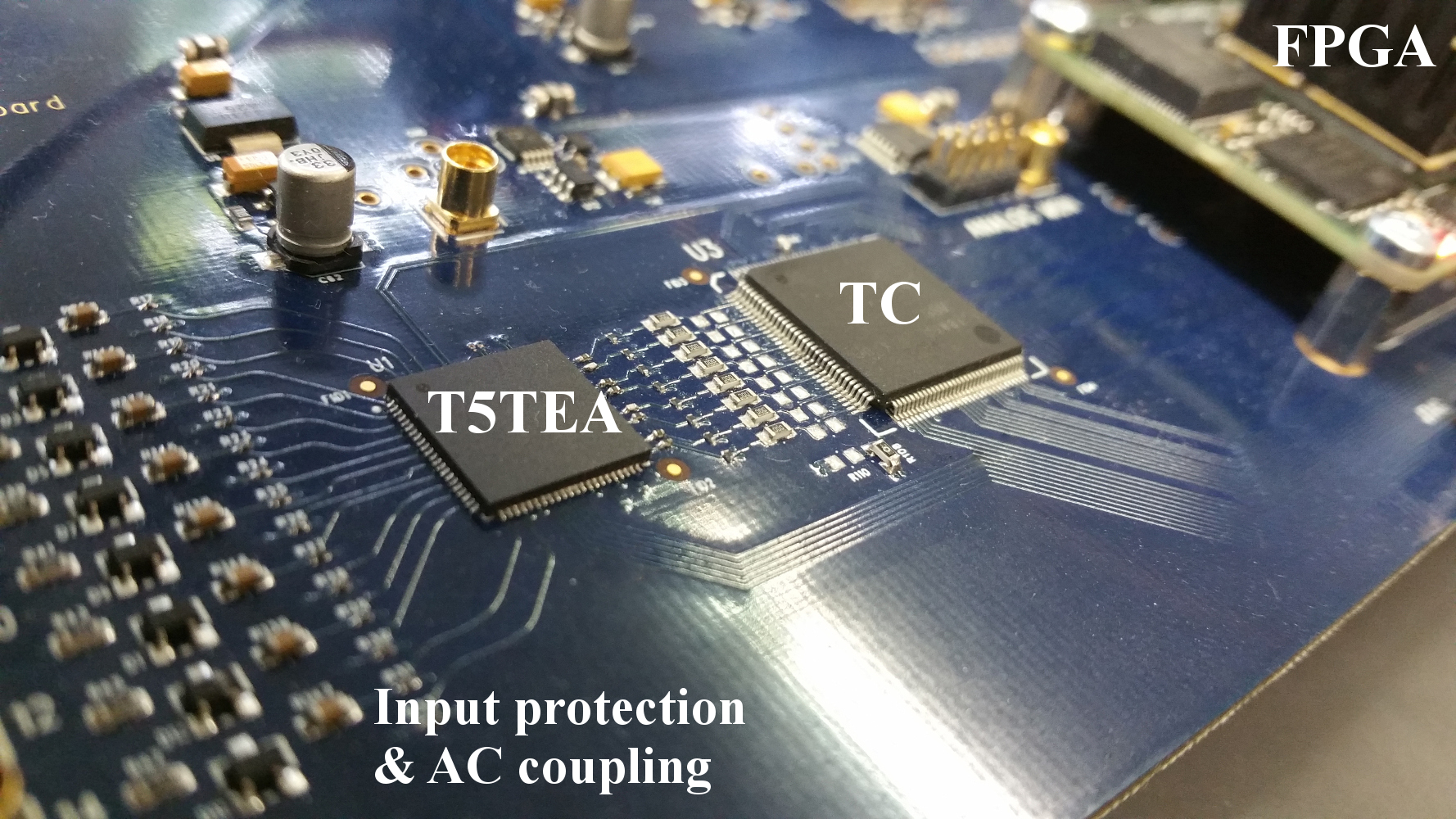}}
  \caption{Evaluation board with the T5TEA and TARGETC ASICs. This board was used for all measurements shown is this paper.}
  \label{fig:evalboard}
\end{figure}

For all the measurements presented in this paper a custom evaluation board was used. This board only contains the two ASICs, TARGET C and T5TEA, and all necessary components to control them, notably a FPGA. Figure~\ref{fig:evalboard} shows a photograph of the evaluation board used for all the measurements.

\newpage
\section{CHARACTERIZATION OF TARGET C AND T5TEA}

\subsection{TARGET C Sampling/Digitization Performance}

The TARGET C ASIC is, for the most part, identical to the sampling and digitization part of TARGET 7 with the exception of an increased digitization clock speed of 500 MHz allowing a faster digitization compared to TARGET 7 and the split off trigger path. 
\begin{wrapfigure}{r}{7.2cm}
\vspace{-0.5cm}
  \centerline{\includegraphics[width=0.5\textwidth]{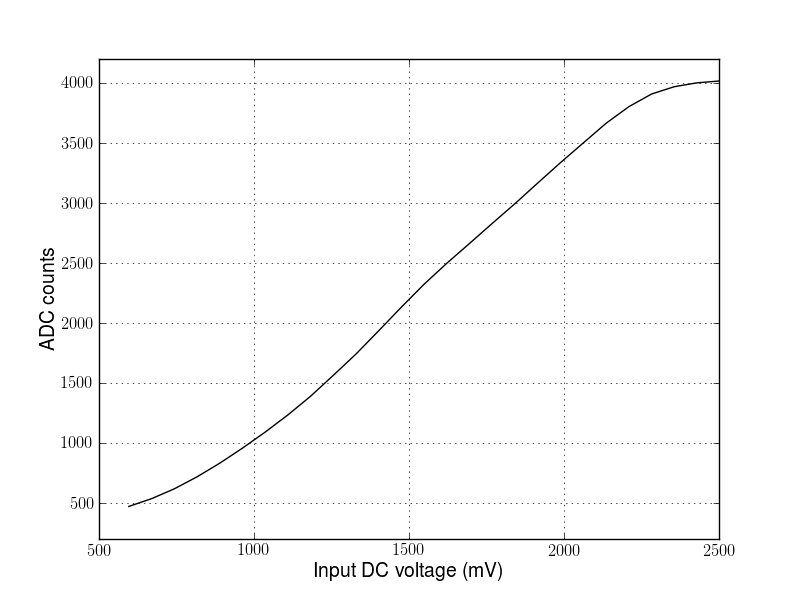}}
  \caption{Transfer function recorded with the onboard DAC at 500 MHz digitization speed.}
  \vspace{-0.2cm}
\end{wrapfigure}
The DC transfer functions and thus the achievable dynamic range, DC noise, and actual digitization time can be controlled with various ASIC parameters set on the fly via the companion FPGA. We tuned the transfer function parameters of TARGET C with the goal of a dynamic range comparable to TARGET 7 with a small integrated non linearity and a smooth transition to the saturated regime. Since the TARGET ASICs record the full waveform for every event a good charge reconstruction is also guaranteed in the saturated regime of the ASIC transfer function using waveform parameters like the rising or falling edge of the pulse. The measured transfer function of TARGET C is shown in Figure 2 and its dynamic range spans 1.9 V with a DC noise of $\sim1$ mV. So, TARGET C shows the same performance as TARGET 7 as it is expected.

The knowledge of the DC transfer function is used to convert the raw ADC counts from the digitizer into a voltage. Examples of calibrated sinusoids and a pulse equivalent to the input to TARGET expected in the camera are shown in Figure 3. The measured sinusoid yields a sampling frequency of $0.9986 \pm 0.0004$ GSa/s which can be corrected to the expected value of 1 GSa/s by further tuning.

\begin{figure}[!htb]
   \begin{minipage}{0.44\textwidth}
   \vspace{-0.44cm}
     \centering
     \includegraphics[width=1.0\linewidth, height=5cm, keepaspectratio]{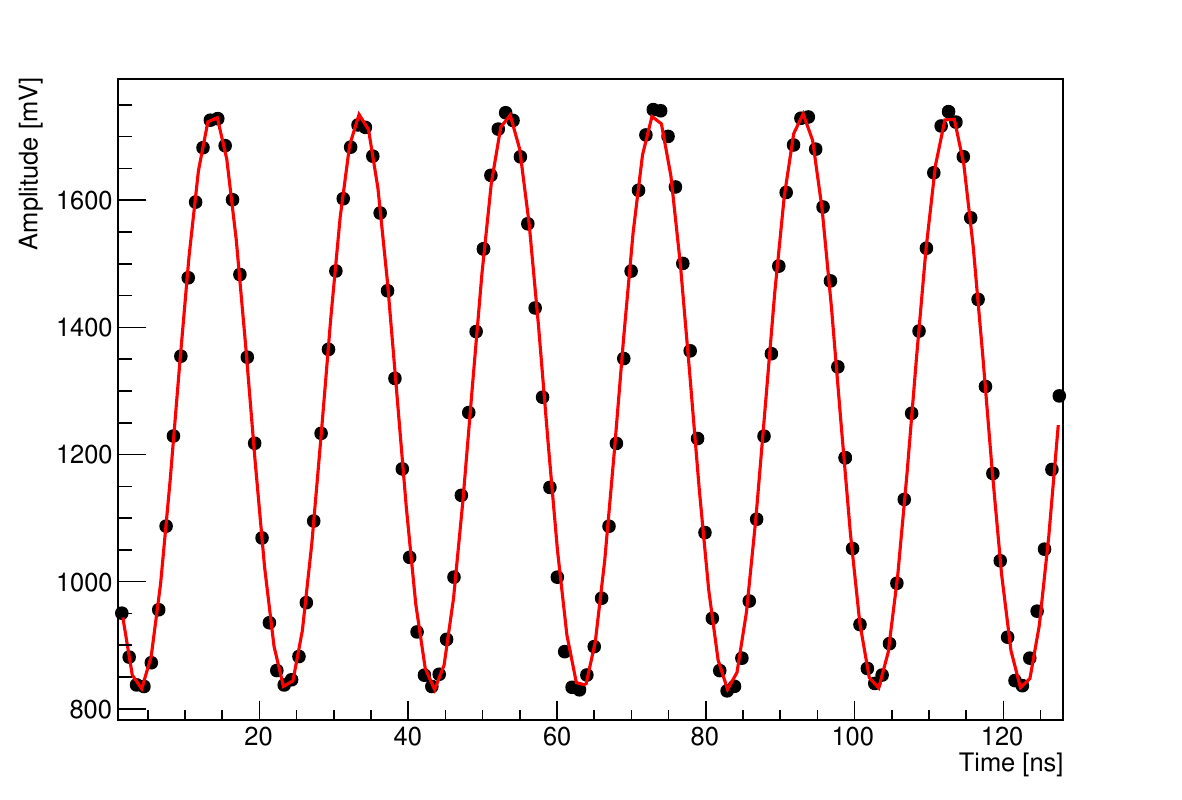}
     \caption*{\small{(a) Calibrated data recorded with the TARGET evaluation board for injected 50 MHz sinusoids.}}\label{Fig:Data1}
   \end{minipage}\hfill
   \begin {minipage}{0.51\textwidth}
     \centering
     \includegraphics[width=1.05\linewidth, height=5.5cm, keepaspectratio]{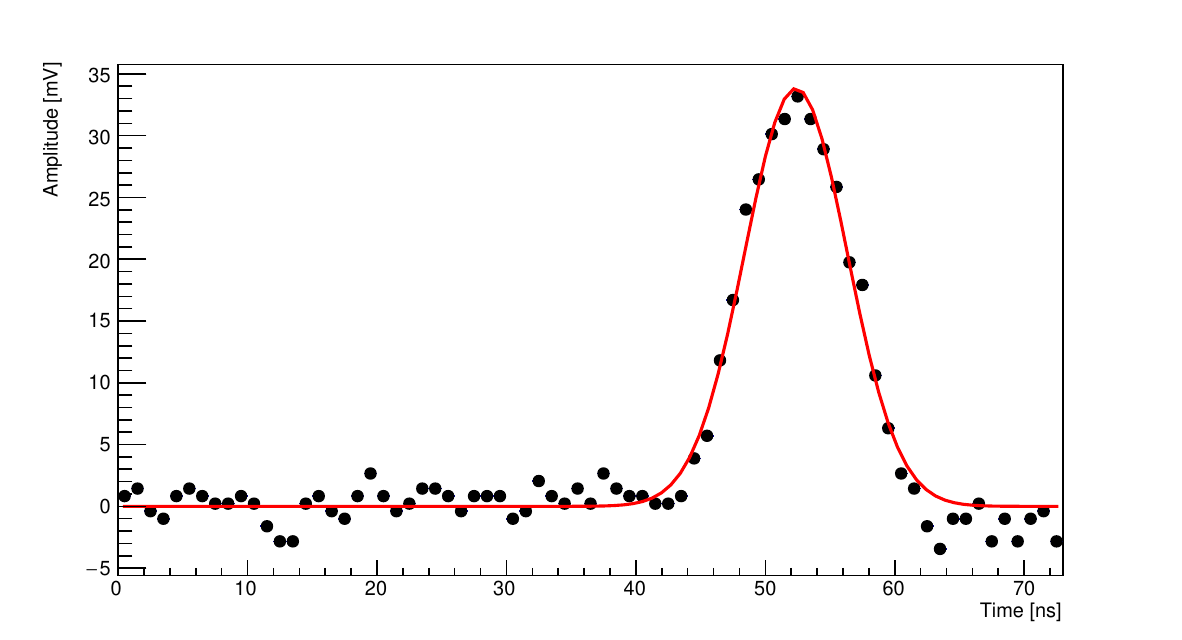}
     \caption*{\small{(b) Calibrated data for  an 35 mV injected pulse with a FWHM width of 10ns. The width is equivalent to the expected input from the camera.}}\label{Fig:Data2}
   \end{minipage}
   \caption{Sinusoid and pulse injected with a function generator. The pulse events were recorded with a trigger rate of 1 kHz.}
\end{figure}

One of the most important performance figures of the TARGET readout electronics is its charge resolution. Here, we expect again a performance as good as in TARGET 7. The split of the sampling/digitization and trigger path might nevertheless have a positive impact on the performance of TARGET C.  The charge resolution measurement shown in Figure 4 was performed using a simple integration method with a window of 16 ns and results in a resolution of $\le4$ \% at 10 pe (assuming a conversion number of 4mV/pe). This is expected to fulfill the camera requirements. However, more complex methods using the full waveform information might yield an even better performance. In the final camera there will be additional noise contributions from the photodetectors and the amplifying/shaping circuits that will degrade the overall charge resolution with respect to TARGET alone.
\vspace{-0.5cm}
\begin{figure}[htbp]
  \centerline{\includegraphics[width=0.42\textwidth]{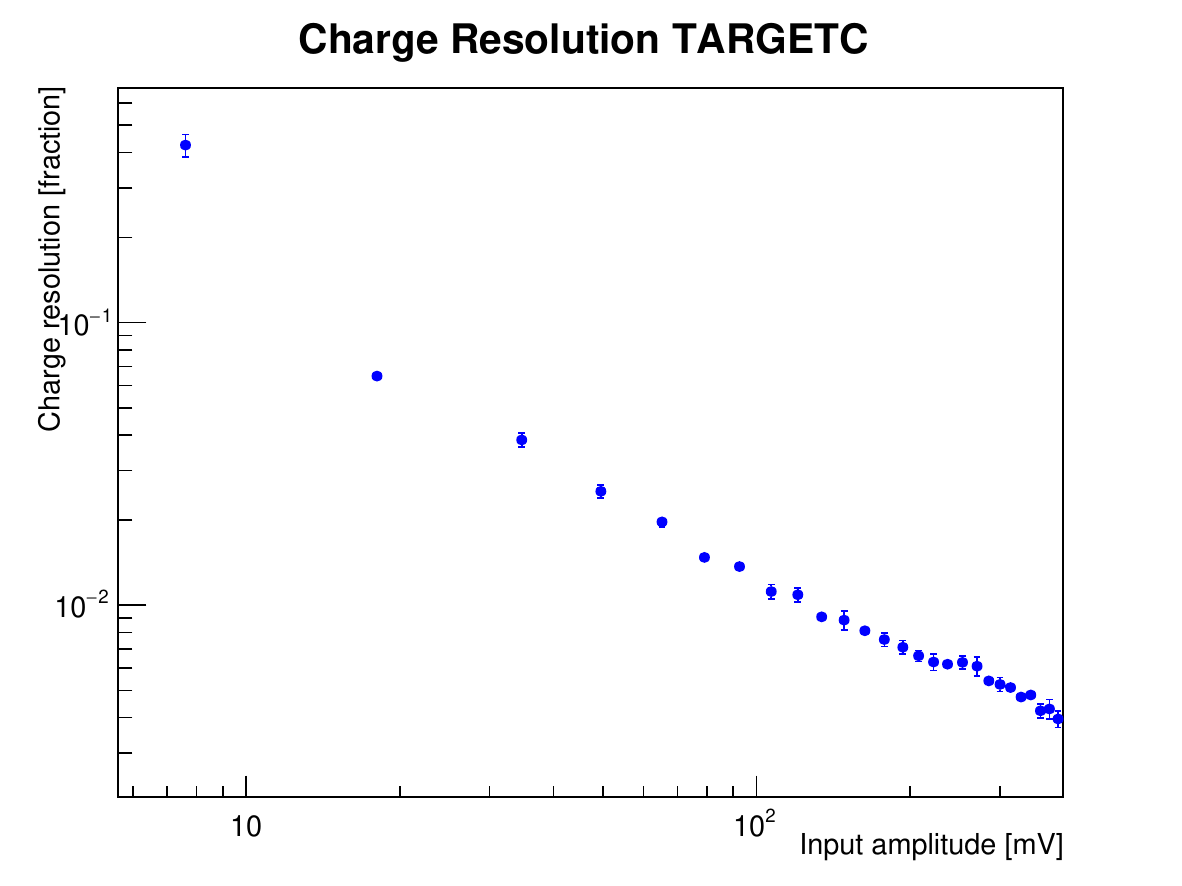}}
  \caption{TARGET C charge resolution for injected function generator pulses. The resolution was determined as ratio of the standard deviation of the reconstructed charge and the mean value. With the expected conversion number of 4mV/pe the TARGET C performance is equal or even better than TARGET 7. }
\end{figure}

\newpage
\subsection{T5TEA Trigger Performance}

The trigger ASIC T5TEA can be controlled by various parameters but the two most important ones, which have the largest impact on the trigger performance, are \texttt{PMTref4} and \texttt{Thresh}. \texttt{Thresh} sets the reference voltage for the comparator with which the amplitude of the input signal is compared and, therefore, directly sets the threshold of the trigger ASIC. On the other hand \texttt{PMTref4} controls the reference voltage for the summing amplifier which performs the analog sum of the four input signals of one trigger group and is able to shift the signal closer to the comparator threshold of \texttt{Thresh}. All parameters are controlled by a 12 bit digital-to-analog converter and have an operating voltage range of 0 to 2.5 V.

For all trigger measurements pulses with varying amplitudes, a full width at half maximum of 10 ns with rise times of 5.4 ns and a frequency of 1 kHz were applied to one channel. The number of occurring trigger signals of one group in a certain time window can be counted by the FPGA. It should be noted that for all trigger measurements sampling of TARGET C was enabled which leads to a larger noise but reflects the environment of normal operations. The efficiency of the trigger can be calculated as $\epsilon = N_{trigger} / N_{pulses}$, where $N_{trigger}$ is the number of trigger signals measured by the FPGA and $N_{pulses}$ the number of injected pulses by the function generator.

\begin{wrapfigure}{r}{6.3cm}
\includegraphics[width=0.42\textwidth]{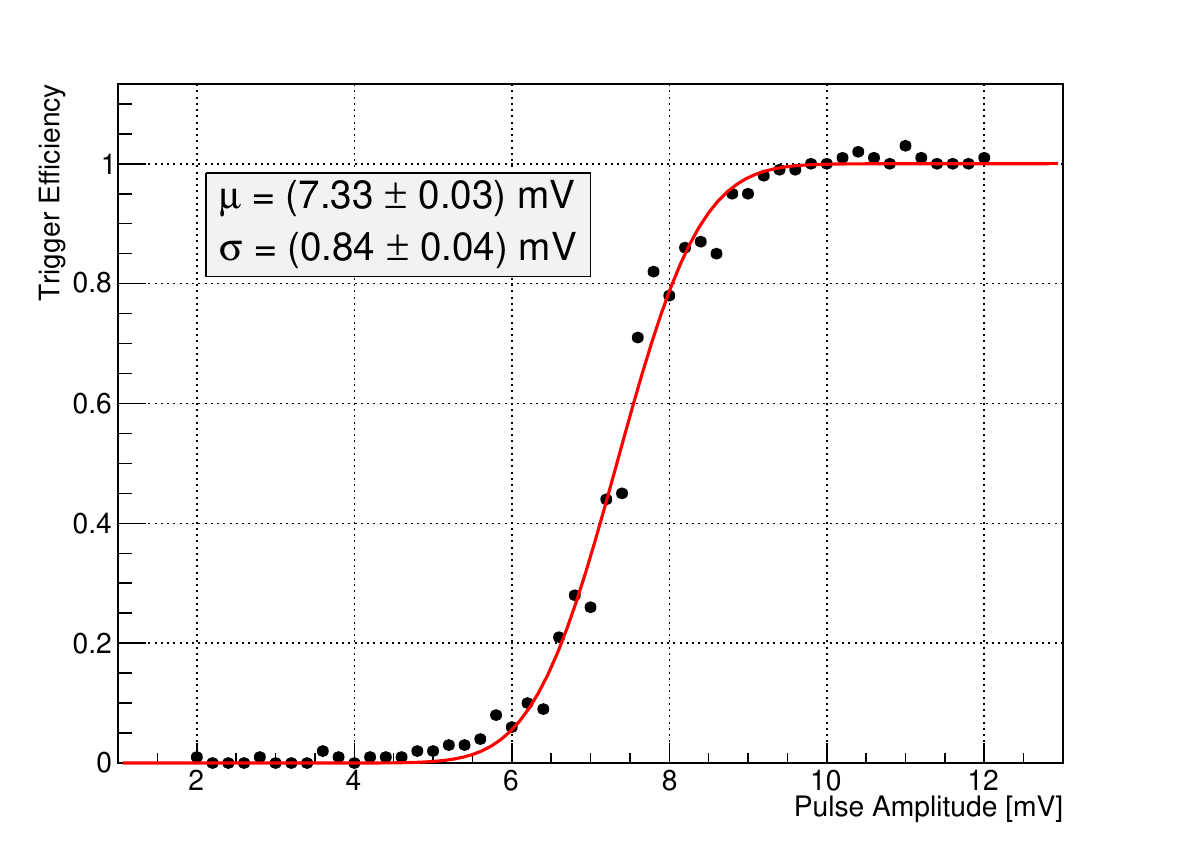}
\caption{Efficiency curve of T5TEA. The black data points show the calculated trigger efficiency as a function of the applied signal amplitude. The red curve is the best-fit Gaussian error function.}
\label{fig:triggereff}
\end{wrapfigure}

To investigate the general performance of the trigger ASIC one can apply signals with varying amplitudes for a given parameter combination of \texttt{PMTref4} and \texttt{Thresh}. The trigger efficiency $\epsilon$ as a function of the input amplitude $A$ is shown in Figure~\ref{fig:triggereff}. Here, a Gaussian error function, also called efficiency curve, 
\begin{wrapfigure}{r}{6.3cm}
\vspace{-0.85cm}
\centerline{\includegraphics[width=0.4\textwidth]{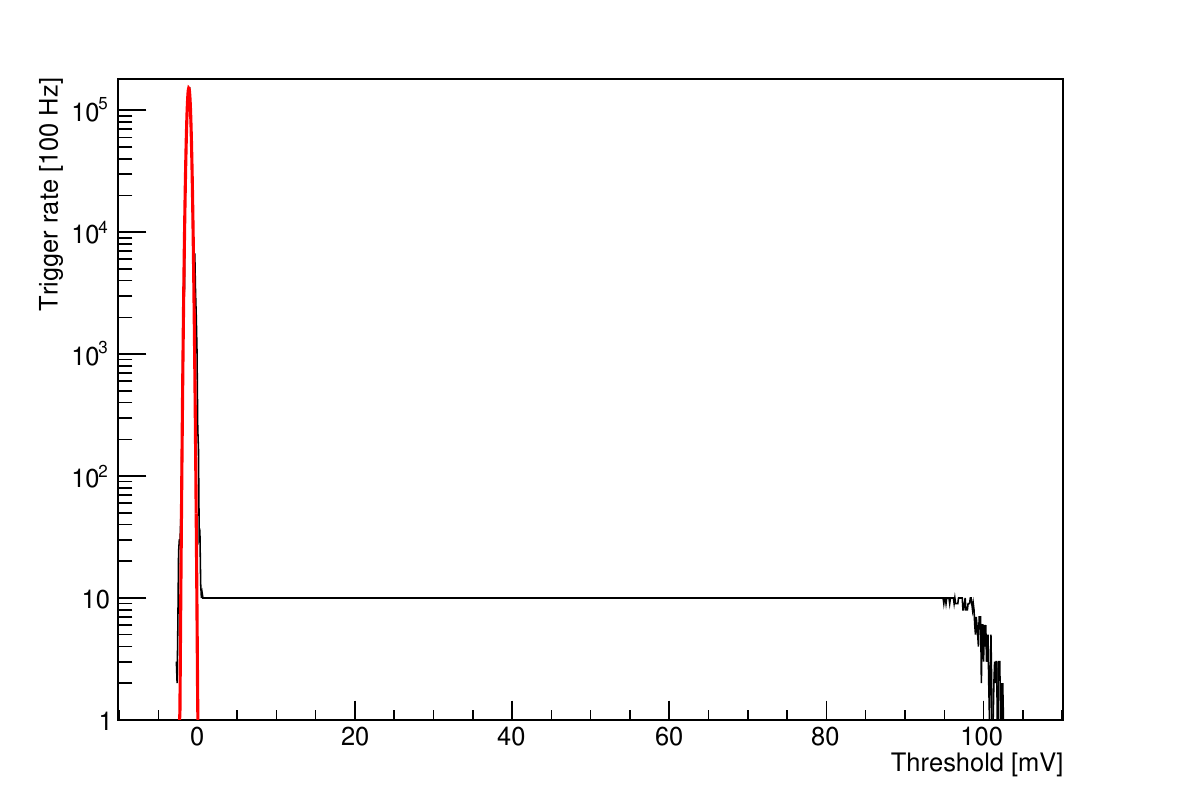}}
\vspace{-0.3cm}
\caption{Trigger rate of T5TEA as a function of the set threshold (\texttt{Thresh}).}
\label{fig:triggerrate}
\vspace{-0.2cm}
\end{wrapfigure}
\begin{equation}
f(A;\mu,\sigma) = \frac{1}{2} \left( 1 + \mathrm{erf} \left( \frac{A - \mu }{\sqrt{2} \sigma} \right) \right)
\end{equation}
was fitted to the data points. The fit parameters $\mu$ corresponds to the threshold of 50\% trigger efficiency and $\sigma$ to the width of the rising edge, called noise. Here, \texttt{PMTref4} was set to 1980 DAC and \texttt{Thresh} was adjusted to result in the lowest threshold $\mu$. With our current evaluation board design and these settings we get a lowest threshold of $\leq 8$ mV (2 pe) and a corresponding noise of $\leq 1$ mV (1/4 pe) which meets the desired camera performance (assuming a conversion number of 4 mV/pe) and represents a significant improvement in comparison to the TARGET 5 \cite{T5} and TARGET 7 \cite{ICRC2015} performance. 

The response to varying \texttt{Thresh} can be best seen in Figure~\ref{fig:triggerrate} where the trigger rate was measured as a function of the set threshold. Pulses with a constant amplitude of 100 mVpp were applied to one channel and \texttt{Thresh} was varied to change the threshold of the trigger ASIC. In this figure one can recognize different domains: on the right side the trigger rate rises to 1 kHz (corresponding to 100\% trigger efficiency) at 100 mV (\texttt{Thresh}: $\sim 2050$ DAC) and stays constant until the threshold crosses the baseline of the signal and begins to trigger at random fluctuations ("baseline noise") at 0 mV (\texttt{Thresh} $\sim 3100$ DAC), which causes the large trigger rate (peak on the left).  

\begin{figure}[htbp]
\begin{minipage}{0.4\textwidth}
\centerline{\includegraphics[width=1\textwidth]{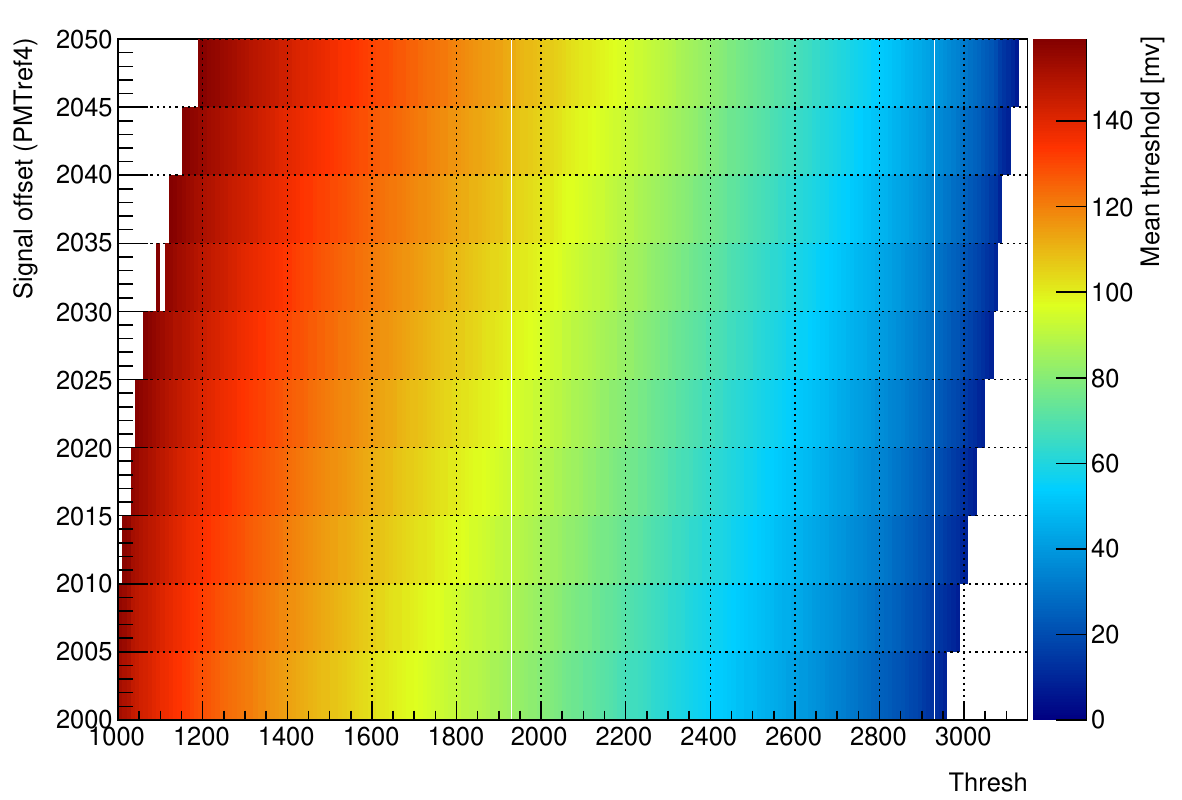}}
\centering\caption*{\small{(a) Trigger threshold}}
\end{minipage}
\begin{minipage}{0.4\textwidth}
\centerline{\includegraphics[width=1\textwidth]{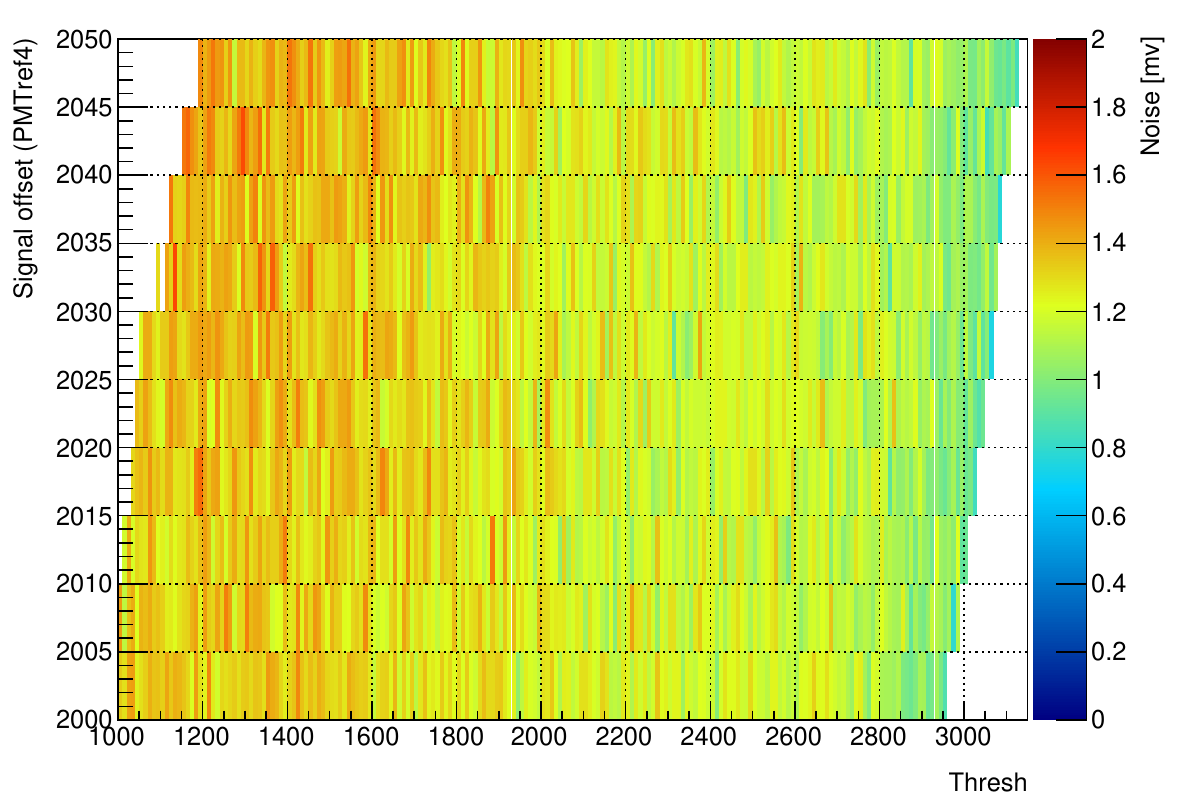}}
\centering\caption*{\small{(b) Noise}}
\end{minipage}
\caption{Scan through the \texttt{PMTref4} - \texttt{Thresh} space. Shown are the results which were determined by measuring efficiency curves. The white areas correspond to a not working trigger (right) and too large thresholds (values $> 150$ mV, left).}
\label{fig:para-scan}
\end{figure}

Also a scan in the parameter space of \texttt{PMTref4} and \texttt{Thresh} was performed: for each combination of the two parameters an efficiency curve was measured and the threshold and noise were determined. The results of the scan are shown in Figure~\ref{fig:para-scan}. The outcome is comparable to the former measurements: over the measured range of \texttt{PMTref4} we get lowest thresholds of $< 10$ mV (2.5 pe) and over the whole range of \texttt{PMTref4} and \texttt{Thresh} we get a noise of $< 2$ mV (0.5 pe), while the noise increases with larger thresholds. The visible fluctuations of the noise are caused by statistical fluctuations. It shows that thresholds between 2.5 pe and 40 pe can be set to 0.5 pe precision. Thus, the desired performance of the trigger ASIC T5TEA is met even at this early stage of our board design.

\newpage
\section{OUTLOOK}
\begin{wrapfigure}{r}{6.5cm}
\vspace{-0.75cm}
\centerline{\includegraphics[width=0.4\textwidth]{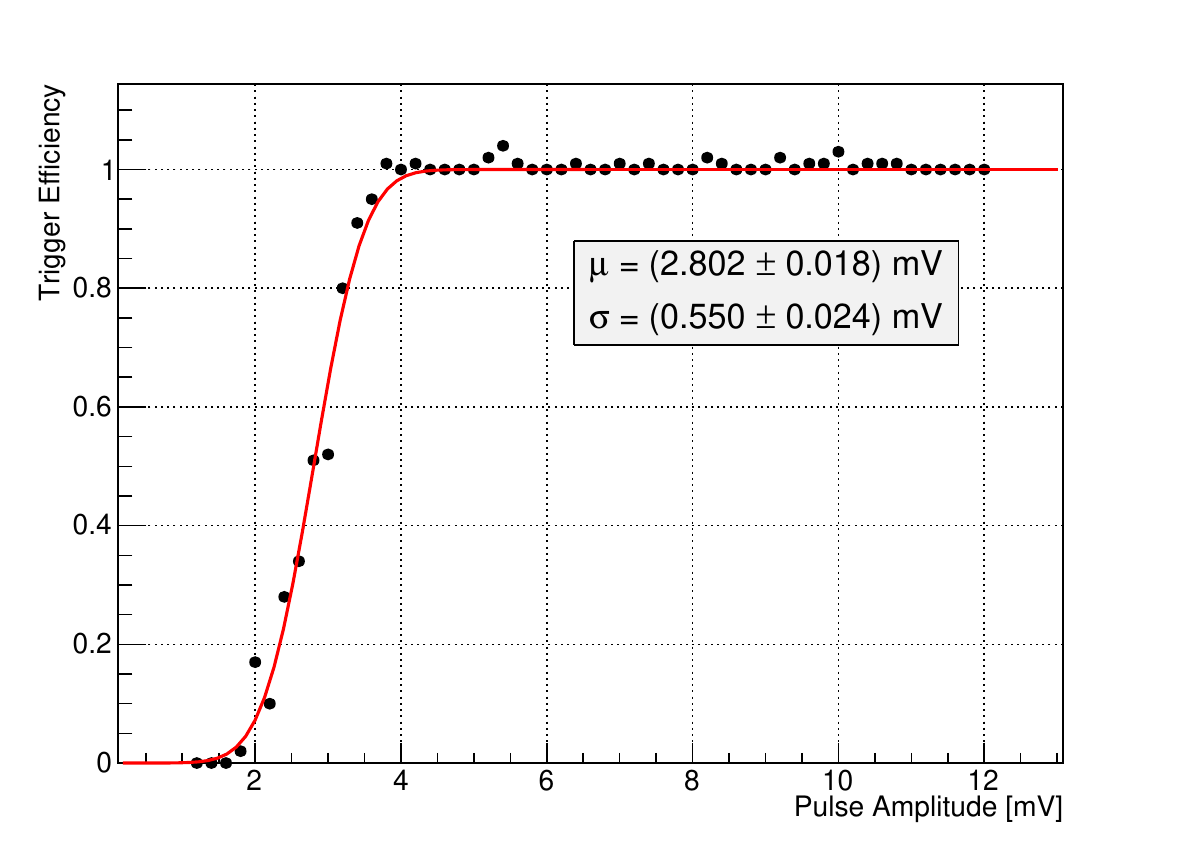}}
\caption{Efficiency curve of T5TEA measured with the newest design of the evaluation board.}
\label{fig:triggereff_new}
\vspace{-0.5cm}
\end{wrapfigure}
Since the presented measurements do not represent final results, further improvements are expected: on the one hand by tuning all available parameters and on the other hand by improving the design of the evaluation boards and camera modules. Figure~\ref{fig:triggereff_new} shows an efficiency curve measured with a debugged and optimized design of our evaluation board where the termination resistors were changed and bypassing capacitors were added. These improvements lowered the threshold to $\le3$ mV ($3/4$ pe) and the noise to $\le0.6$ mV (0.15 pe). Currently ongoing investigations address the whole readout chain with SiPM, buffer and shaper board and an evaluation board. The next steps in the near future include production and testing of complete TARGET modules and the production of an entire camera, called CHEC-S, which will be used in the prototype GCT telescope.

\ \\
\section{ACKNOWLEDGMENTS}
We gratefully acknowledge support from the agencies and organizations under Funding Agencies at \url{www.cta-observatory.org}.


\bibliographystyle{aipnum-cp}%
\bibliography{target}%

\end{document}